# OPTIMIZATION OF MULTIPLE VEHICLE ROUTING PROBLEMS USING APPROXIMATION ALGORITHMS

R. Nallusamy[1*], K. Duraiswamy[2], R. Dhanalaksmi[3] and P. Parthiban[4]

[1,2] Department of Computer Science and Engineering, K.S.Rangasamy College of Technology,
Tiruchengode-637215, India

[*]E-mail: nallsam@rediffmail.com

[3]D-Link India Ltd, Bangalore, India

[4]Department of Production Engineering, National Institute of Technology Tiruchirappalli, India

## ABSTRACT

This paper deals with generating of an optimized route for multiple Vehicle routing Problems (mVRP). We used a methodology of clustering the given cities depending upon the number of vehicles and each cluster is allotted to a vehicle. k- Means clustering algorithm has been used for easy clustering of the cities. In this way the mVRP has been converted into VRP which is simple in computation compared to mVRP. After clustering, an optimized route is generated for each vehicle in its allotted cluster. Once the clustering had been done and after the cities were allocated to the various vehicles, each cluster/tour was taken as an individual Vehicle Routing problem and the steps of Genetic Algorithm were applied to the cluster and iterated to obtain the most optimal value of the distance after convergence takes place. After the application of the various heuristic techniques, it was found that the Genetic algorithm gave a better result and a more optimal tour for mVRPs in short computational time than other Algorithms due to the extensive search and constructive nature of the algorithm.

## 1. INTRODUCTION

Problems of combinatorial optimization are characterized by their well-structured problem definition as well as by their huge number of action alternatives in practical application areas of reasonable size [9]. Utilizing classical methods of Operations Research often fails due to the exponentially growing computational effort. Therefore, in practice heuristics and meta-heuristics are commonly used even if they are unable to guarantee an optimal solution.

### 1.1 Scope and objectives of research

A careful analysis of literature on the variants and methodologies of combinatorial optimization problems reveals that some of the variants of combinatorial optimization problems are yet to be explored to solve using meta-heuristics techniques [9],[10]. These include:

- Multiple Vehicle Routing Problem (mVRP) and multiple Traveling Salesman Problem (mVRP)
- mVRP with balanced allocation of nodes with single objective or multiple objectives

Many of the authors [2] have suggested the use of a constructive heuristic to obtain good initial solutions for a meta-heuristic so that its convergence can be accelerated. Only a few authors have considered the use of hybrid approaches to solve different variants of combinatorial optimization problems.

### 1.2 Vehicle routing problem

In VRP a number of cities have to be visited by a vehicle which must return to the same city where it started. In solving the problem one tries to construct the route so that the total distance traveled is minimized. Every vehicle starts from the same city, called depot and must return at the end of its journey to this city again.

If n is the number of cities to be visited then (n-1)! is the total number of possible routes. As the amount of input data increases the problem increases in complexity, thus the computational time needed renders this method impractical for all but a smaller number of cities. Rather than considering all possible tours, heuristic algorithms for solving the VRP are capable of substantially reducing the number of tours to be taken into consideration.

### 1.3 Multiple vehicle routing problem

A generalization of the well-known vehicle routing Problem is the multiple vehicle routing problem, which consists of determining a set of routes for m vehicles. The mVRP can in general be defined as follows: Given a set of nodes, let there be m vehicle located at a single depot node. The remaining nodes (cities) that are to be visited are called intermediate nodes. Then, the mVRP consists of finding tours for all m vehicles, which all start and end at the depot, such that each intermediate node is visited exactly once and the total cost of visiting all nodes is minimized.

## 2. REVIEW OF EXISTING WORK

Many methods have been suggested for obtaining optimized route[2,3]. Rizzoli *et al*.[1] have focused on the Application of Ant Colony Optimization on the Vehicle Routing Problem and its real world application. Potvin[2] has worked on the survey of the genetic algorithms in his study he has given simple genetic algorithms and various extensions for solving Traveling Salesman Problem (TSP). He has worked both on the random and the classical problems [6].

Schabauer, Schikuta, and Weishaupl [3] have worked on to solve traveling salesman problem heuristically by the parallelization of self-organizing maps on cluster architectures. Allan Larsen has worked on the dynamic factors of vehicle routing problem. He has investigated the dynamics of the vehicle routing problem in order to improve the performances of existing algorithms and as well as developed new algorithms [4]. Jorg Homberger and Hermann Gehring have worked on vehicle routing problems on time windows. In this they have designed an optimal set of routes that will service the entire customer with constrains being taken care of properly. Their objective function minimizes both the total distance traveled and the number of salesmen being used[5].

Al-Dulaimi and Ali [6] have proposed a software system to determine the optimal route of the traveling salesman using Genetic Algorithm (GA). The system proposed starts from a matrix of the calculated Euclidean distances to the cities to be visited by the salesman. The new generations are formed from this until proper path is obtained. Chao, Ye and Miao [7] have developed a two level genetic algorithm which favors neither intra-cluster path or inter-cluster path. The results from the study indicate that the algorithm proposed is more effective than the existing algorithms.

A.E. Carter, C.T. Ragsdale have developed a new approach to solve mTSP. The method proposes new set of chromosomes and related operators for the mTSP and compares theoretical properties and computational performance of the proposed technique. The computational technique shows that the newer technique results in the smaller search space and produces better solutions [8].

Mitrovic-Minic and Krishnamurti[7] have worked onto to find the lower and upper bound required for the number of vehicles to serve all locations for multiple traveling Salesman problem with time windows. They have introduced two types of precedence graphs namely the start time precedence graphs and the end time precedence graphs. The bounds are generated by covering the precedence graphs with minimum number of paths. The bounds which are tight and loose are compared and the closeness of such instances were discussed.

Researchers on the VRP have proved that the VRP is a NP-complete combinatorial optimization problem. They have theorized that if an algorithm is guaranteed to find the optimal solution in a polynomial time for the VRP, then efficient algorithms could also be found for all the other NP-complete problems[10]-[15].

### 1.4 Research gap and proposed work

From the review, we understood that most of the problems involved solving the conventional vehicle routing problem or traveling salesman problem using exact as well as meta-heuristic methods for solving the same. They however scarcely dealt with the multiple vehicle routing Problem which represents the realistic case of more than one vehicle. To the best of our knowledge, from the literature review, no efficient meta-heuristic algorithms exist for the solution of large-scale mVRPs. Also, the solution procedures based on transforming the mVRP to the standard VRP do not seem efficient, since the resulting VRP is highly degenerate, especially with the increasing number of vehicles. Hence, an analysis is made and an heuristic is formed to transform mVRP to VRP and to optimize the tour of an individual. We decided to deal with the less frequently approached and more realistic multiple vehicle routing problem along with a specialized clustering heuristic, namely k-means clustering algorithm

## 3. PROBLEM BACKGROUND AND PROBLEM FORMULATION

The mathematical structure of the VRP is a graph where the cities are the nodes of the graph. Connections between pairs of cities are called edges and each edge has a cost associated with it which can be distance, time or other attribute. If n is the input number of vertices representing cities, for a weighted graph G, the VRP problem is to find the cycle of minimum costs that visit each of the vertices of G exactly once.

There are many mathematical formulations for the VRP, employing a variety of constraints that enforce the requirements of the problem. The following notation is used: n-The number of cities to be visited; the number of nodes in the network; i, j, k- Indices of cities that can take integer values from 1 to n; t-The time period, or step in the route between the cities; $x_{ijt}$-1 if the edge of the network from i to j is used in step t of the route and 0 otherwise; $d_{ij}$-The distance or cost from city i to city j. The following is an example of one linear programming formulations of the VRP problem:

The objective function (Z) is to minimize the sum of all costs (distances) of all of the selected elements of the tour:

$$Z = \sum_{i=1}^{n}\sum_{j=i}^{n}\sum_{t=1}^{n} d_{ij} x_{ijt} \quad (1)$$

The tour is subject to the following constraints. For all values of t, exactly one arc must be traversed, hence:

$$\sum_{i}\sum_{j} x_{ijt} = 1 \text{ for all } t \quad (2)$$

For all cities, there is just one other city which is being reached from it, at some time, hence:

$$\sum_{j}\sum_{t} x_{ijt} = 1 \text{ for all } i \quad (3)$$

For all cities, there is some other city from which it is being reached, at some time, hence

$$\sum_{i}\sum_{t} x_{ijt} = 1 \text{ for all } j \quad (4)$$

When a city is reached at time t, it must be left at time t+1, in order to exclude disconnected sub-tours that would otherwise meet all of the above constraints. These sub-tour elimination constraints are formulated as:

$$\sum_{i} x_{ijt} = \sum_{k} x_{jkt+1} \text{ for all j and t} \quad (5)$$

In addition to the above constraints the decision variables are constrained to be integer values in the range of 0-1:

$$0 \leq x_{ijt} \leq 1 \quad (6)$$

## 4. MATERIALS AND METHODS

### 4.1. Assumptions

All the salespersons have to start from a common depot and after traveling through a set of cities, they should return back to the starting depot. There are no capacity constraints and no cost constraints. But, all the cities must be visited by any one of the salesperson and each salesperson has to visit a particular city exactly once.

### 4.2. Transformation of mVRP to VRP

The search space for the solution increase as the number of cities decreases and vice-versa. If there are N cities then the search space will be N! and the computational time also high accordingly. Hence to reduce the burden of mathematical complexity N value should be reduced and this is achieved by clustering. The following heuristics were used for solving the given 180 cities 6 vehicles problem. City number 100 is considered to be the headquarters of all the vehicles.

### 4.3. k-means clustering

Simply speaking k-means clustering is an algorithm to classify or to group the objects based on attributes/features into k number of group. k is a positive integer number. The grouping is done by minimizing the sum of squares of distances between data and the corresponding cluster centroid [2].

The main idea is to define k centroids, one for each cluster. These centroids should be placed in a cunning way because of different location causes different result. So, the better choice is to place them as much as possible far away from each other. The next step is to take each point belonging to a given data set and associate it to the nearest centroid.

When no point is pending, the first step is completed and an early groupage is done. At this point we need to re-calculate k new centroids as barycenters of the clusters resulting from the previous step. After we have these k new centroids, a new binding has to be done between the same data set points and the nearest new centroid. A loop has been generated. As a result of this loop we may notice that the k centroids change their location step by step until no more changes are done. In other words centroids do not move any more.

This produces a separation of the objects into groups from which the metric to be minimized can be calculated. Although it can be proved that the procedure will always terminate, the k-means algorithm does not necessarily find the most optimal configuration, corresponding to the global objective function minimum. The algorithm is also significantly sensitive to the initial randomly selected cluster centers. The k-means algorithm can be run multiple times to reduce this effect.

The algorithm is composed of the following steps:

- *Place k points into the space represented by the objects that are being clustered. These points represent initial group centroids*
- *Assign each object to the group that has the closest centroid*
- *When all objects have been assigned, recalculate the positions of the K centroids*
- *Repeat steps 2 and 3 until the centroids no longer move. This produces a separation of the objects into groups from which the metric to be minimized can be calculated*

### 4.4. Application of GA to the given mVRP

Genetic algorithms emulate the mechanics of natural selection by a process of randomized data exchange. The fact that they are able to search in a randomized, yet directed manner, allows them to reproduce some of the innovative capabilities of natural systems. GAs work by generating a population of numeric vectors called chromosomes, each representing a possible solution to a problem. The individual components within a chromosome are called genes. New chromosomes are created by crossover or mutation. Chromosomes are then evaluated according to a fitness function, with the fittest surviving and the less fit being eliminated. The result is a gene pool that evolves over time to produce better and better solutions to a problem. The GAs search process typically continues until a pre-specified fitness value is reached, a set amount of computing time passes or until no significant improvement occurs in the population for a given number of iterations. The key to find a good solution using a GA lies in developing a good chromosome representation of solutions to the problem.

### 4.5. Algorithm for genetic algorithm

*P = Generate Initial Population of Solutions;*
*While (stopping criterion not met)*
    *For (X $\in$ P) C(X) = Evaluate Cost of X;*
        *P′ = Select Fittest Individuals from P to Form Mating Pool;*
        *P″ = $\varnothing$;*
        *Repeat (until enough children produced)*
            *Select X1 and X2 at Random From P′;*
            *Apply Mating Procedures to X1 and X2 to Produce Xchild;*
            *P″ = P″ $\cup$ Xchild;*
        *End Repeat;*
        *For (X $\in$ P″)*
        *Apply Random Mutation to X;*
    *End For;*
    *P = P″;*
*End While;*
*Output C(X) where X is fittest individual in P;*
*End.*

There is an optimal set of cities allocated to every vehicle after performing the k-means clustering algorithm. Genetic algorithm is now applied for every such cluster of cities and iteration is performed several times to find an optimal value for the distance traveled by each vehicle.

### 4.6. Formulation of the fitness value

A fitness function is a particular type of objective function that quantifies the optimality of a solution so that that particular chromosome may be ranked against all the other chromosomes. Optimal chromosomes, or at least chromosomes which are more optimal, are allowed to breed and mix their datasets by any of several techniques, producing a new generation that will be even better. The shorter the route, the higher is the fitness value. Hence, we formulated the fitness value or function as the inverse of the reciprocal of the distance traveled in each sequence. Hence, fitness value or function $f = 1/d_i$, where $d_i$ is the distance traveled by a vehicle after covering all the cities allocated to him.

### 4.7. Selection of the initial population

After optimally assigning a definite group of cities to a vehicle using the clustering meta-heuristic, a group of 10 sequences of all the cities in the cluster are selected from the universal population of all possible sequences. This is performed randomly by using probabilistic selection based on the favorable fitness value.

### 4.8. Crossover and Cross over probability

The population is arranged in descending order of the sequence's fitness value. The top 7 members of the list are selected and random values are assigned to each chromosome using random number generation function. In our problem we took the crossover probability as 0.8. Simple chromosome 'crossover' is chosen for our reproduction scheme because it is the simplest. Finally, we generate an initial population of 200 random chromosomes and run the program. We are astounded as the program produces nothing of use.

From the review we identified that the order in which data is encoded in the chromosome is important for VRP and that a simple crossover reproduction mechanism is not suitable in these circumstances. With this in mind, we need a scheme analogous to simple crossover, but one which preserves the solution viability while allowing the exchange of ordering information. One such scheme is Partially Matched Crossover. In this scheme, a crossing region is chosen by selecting two crossing sites. This type of partial matched crossover is done for every chromosome with the next best to yield an offspring. The offspring replaces the parent chromosome if the fitness value for the former is higher than that of the later.

### 4.9. Mutation probability

From the population of the chromosomes modified after crossover, a set of chromosomes are selected for mutation based on the mutation probability. In our problem we took the mutation probability as 0.1. After performing the mutation on the selected chromosomes, the next set of 10 chromosomes is taken as the initial population for the next iteration. The previous crossover and mutation steps are repeated for several iterations till the fitness value of the best chromosome in a given population converges to a constant value. This yields the result for the optimal distance traveled by one vehicle.

### 5. IMPLEMENTATION, RESULTS AND DISCUSSION

The solution to the problem is attained using two-stage heuristics. The first-stage involves the conversion of a mVRP to VRP using k-means Clustering algorithm. Even though we get, the cities allocated to a vehicle, it is important to generate a tour and improve it. Second-stage is meta-heuristic approach comprising GA to optimize the tour for m vehicles. Then the effective solution generated by GA is studied and is compared with the results obtained from other methods. This gave an advantage of choosing a consistent approach to particular types of problems. The problem was implemented in MATLAB 7.0 with Pentium IV processor system.

Here we used randomly generated coordinates of 180 cities within 35X35 square units space. After performing k-means clustering algorithm, GA is applied. Figure 1 and 2 show the convergence of results in GA. Table 1 shows the results of k-means clustering. Table 2 shows the results of GA with other algorithms. Algorithm was applied to the given problem iteratively; we found that optimal results of the distance were obtained after performing approximately 300 iterations.

Table 1. Cities allocated to the 6 vehicles

| Cluster /vehicle | Number of Cities allocated |
|---|---|
| 1 | 44 |
| 2 | 42 |
| 3 | 25 |
| 4 | 25 |
| 5 | 7 |
| 6 | 37 |

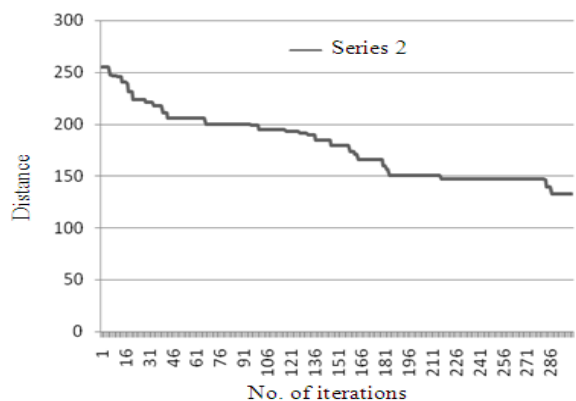
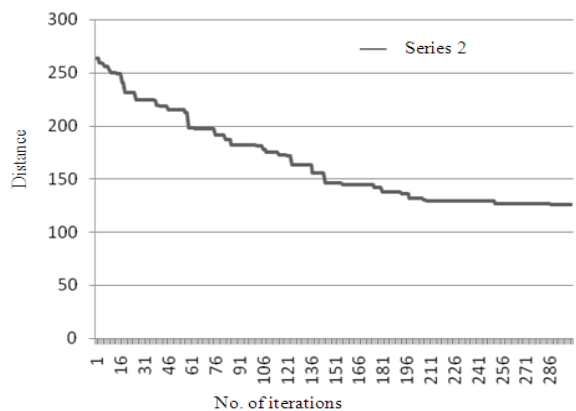

Fig. 1. Convergence diagram of GA for vehicle 1        Fig. 2. Convergence diagram of GA for vehicle 2

Table-2 Results of GA vs other algorithms

| Vehicle | Tabu search (units) | Simulated annealing (units) | GA distance (Units) |
|---|---|---|---|
| 1 | 146.2028 | 138.8827 | 132.393 |
| 2 | 141.643 | 133.4244 | 126.258 |
| 3 | 58.0388 | 52.098 | 51.3596 |
| 4 | 44.8751 | 41.8924 | 41.6981 |
| 5 | 11.6923 | 10.3178 | 10.3284 |
| 6 | 124.6382 | 120.4332 | 115.618 |

## 6. CONCLUSION AND FUTURE SCOPE

From the results obtained, we find that k-means clustering proved to be effective as it was able to group the cities into clusters in an optimal manner and convergence took place in a short execution time and the optimal clusters were obtained. It can be clearly inferred that GA yields better solutions to the mVRP. This result might have been obtained in a lesser computational time than the exact methods like Branch and Bound, Branch and Cut and Cut and Solve techniques. It is not the global optimal solution and might turn out to be a close to global optima or a local optima as the convergence graph showed signs of fall even after a long period of convergence. Moreover, the solution may further fall in value after a large number of iterations. This may cost us further computational time and usage of computer memory to further genetically modify the chromosomes and give a better solution. Hence, Genetic Algorithm makes a compromise between the efficiency and the optimality of the final result obtained. The work can be further extended by balancing the workloads of the salesmen by manipulating between clusters and reducing the standard deviation of the distance values. This will help in improving workers morale.